# Deep Learning Analysis of Cardiac MRI in Legacy Datasets: Multi-Ethnic Study of Atherosclerosis

**Avan Suinesiaputra[1,2], Charlene A Mauger[1], Bharath Ambale-Venkatesh[3], David A Bluemke[4], Josefine Dam Gade[5], Kathleen Gilbert[6], Mark Janse[7], Line Sofie Hald[5], Conrad Werkhoven[6], Colin Wu[3], Joao A Lima[3], Alistair A Young[1,2]**

[1] Department of Anatomy and Medical Imaging, University of Auckland, Auckland, New Zealand

[2] Department of Biomedical Engineering, School of Biomedical Engineering and Imaging Sciences, King's College London, UK

[3] Johns Hopkins Medical Center, Baltimore, USA

[4] Department of Radiology, University of Wisconsin School of Medicine and Public Health, Wisconsin, USA

[5] Department of Biomedical Engineering & Informatics, School of Medicine and Health, Aalborg University, Denmark.

[6] Auckland Bioengineering Institute, University of Auckland, Auckland, New Zealand

[7] Department of Electrical Engineering, Eindhoven University of Technology, Eindhoven, The Netherlands

## Abstract

The shape and motion of the heart provide essential clues to understanding the mechanisms of cardiovascular disease. With the advent of large-scale cardiac imaging data, statistical atlases become a powerful tool to provide automated and precise quantification of the status of patient-specific heart geometry with respect to reference populations. The Multi-Ethnic Study of Atherosclerosis (MESA), begun in 2000, was the first large cohort study to incorporate cardiovascular MRI in over 5000 participants, and there is now a wealth of follow-up data over 20 years. Building a machine learning based automated analysis is necessary to extract the additional imaging information necessary for expanding original manual analyses. However, machine learning tools trained on MRI datasets with different pulse sequences fail on such legacy datasets. Here, we describe an automated atlas construction pipeline using deep learning methods applied to the legacy cardiac MRI data in MESA. For detection of anatomical cardiac landmark points, a modified VGGNet convolutional neural network architecture was used in conjunction with a transfer learning sequence between two-chamber, four-chamber, and short-axis MRI views. A U-Net architecture was used for detection of the endocardial and epicardial boundaries in short axis images. Both network architectures resulted in good segmentation and landmark detection accuracies compared with inter-observer variations. Statistical relationships with common risk factors were similar between atlases derived from automated vs manual annotations. The automated atlas can be employed in future studies to examine the relationships between cardiac morphology and future events.

# 1 Introduction

Cardiovascular magnetic resonance (CMR) is widely used for the non-invasive assessment of cardiac function, and has excellent accuracy and reproducibility for clinical evaluation of cardiac mass and volume (1). The ability of CMR to evaluate all regions of the heart with high signal to noise ratio without harmful radiation exposure has led to its use in several large cohort studies investigating the development of cardiac disease in general populations, including the Multi-Ethnic Study of Atherosclerosis (MESA) (2) and the UK Biobank (3). MESA was the first large epidemiological study to utilize CMR to evaluate pre-clinical characteristics of participants before the onset of clinical symptoms of cardiovascular disease (CVD). The baseline MESA CMR exam was performed between 2000 and 2002 using the common imaging method prevalent at that time: gradient echo cine imaging. However, this imaging method has been largely replaced by steady-state free precession cine imaging in subsequent studies and in clinical practice (4). Due to differences in fundamental properties that comprise image contrast as well as spatial resolution (5), image analysis tools designed for modern steady-state free precession images are likely to have poor performance when applied to 20 year old gradient echo imaging.

Three-dimensional (3D) atlas-based analysis methods have been developed to quantify subtle differences in heart shape (remodelling) and function associated with CVD risk factors such as hypertension, smoking and diabetes (6–10). To date, these methods have only been applied to a limited subset of MESA cases, due to the need for additional image analysis which was not performed as part of the original CMR analysis. This is a recurring problem in large cohort legacy datasets, since a limited amount of annotations are available and manual analysis is unfeasible due to time and resource constraints. A fully automated processing pipeline is therefore necessary to enable more comprehensive analysis and make better use of the large amount of image data acquired.

Deep learning methods, particularly convolutional neural networks (CNN), have demonstrated high accuracy and reproducibility for fully automated image analysis when sufficient training images and high computational power is available (11,12). CNN can automatically learn optimal weights for convolutional operations in each layer to extract image features. It has been applied and adapted for image classification, object recognition, segmentation, and image registration. However, CNN solutions trained on modern steady-state free precession images fail when applied to the MESA dataset. In this study, we developed a pipeline combining deep learning methods with atlas-based analysis to derive a detailed analysis of left ventricular (LV) shape and function from the MESA baseline CMR examination. CNN methods were used to detect anatomical landmarks and left ventricular borders from the MESA gradient echo CMR images and to segment the myocardium. We demonstrate that deep learning networks provide robust and consistent contours and landmarks compared with manual annotations, and that the resulting LV atlas gives similar relationships with CVD risk factors as an atlas generated from manual annotations.

# 2 Methods

## 2.1 Dataset

The MESA study has been described previously (2). Briefly, the baseline CMR exam consisted of 5,098 participants who were initially free from clinically recognized CVD at the time of enrollment (13). Of these 5,003 had adequate MRI data for analysis (Table 1). Of the 5,003 cases, 2,529 (50.5%) were available in the Cardiac Atlas Project (14) and were used to train and validate the deep learning algorithms using ground truth labels from expert-drawn contours and anatomical landmark placements (Training sub-cohort). The remaining 2,474 cases (49.5%) were used as an independent testing cohort to validate the atlas generation pipeline (Atlas Validation sub-cohort). Participant demographics are shown in Table 1, including sub-cohorts used for the training datasets and atlas validation.

CMR images were acquired with 1.5T MR scanners at six different institutions across the United States using Siemens and General Electric scanners between July 2000 and July 2002. Each examination included short- and long-axis cine images, phase-contrast images of the aorta, and black blood aorta images. All images were acquired during breath-holding at resting lung volume. The cine MR images included 10-12 short-axis slices (SAX), single four-chamber (4CH) and single two-chamber (2CH) long-axis (LAX) views using gradient echo imaging, with typical parameters of slice thickness 6 mm, 4 mm gap, field of view 360-400 mm, 256x160 image matrix (smallest 192x160), flip angle 20°, echo time 3-5 ms, repetition time 8-10 ms with 20-30 frames per slice (temporal resolution <50 ms) and pixel size from 1.4 to 2.5 mm/pixel depending on patient size. All participants gave informed consent, and the institutional review board at each site approved the study protocol.

## 2.2 Cardiac MRI analysis pipeline

The overall pipeline is given in Figure 1, combining deep learning networks with atlas-based analysis. Deep learning networks were trained to provide necessary inputs for the customization of the LV shape model, i.e., cardiac landmark points and boundaries of the myocardium. Two different deep learning architectures were used: a VGGNet based network (15) for landmark detection and a U-Net based network (16) for myocardial segmentation. Two types of landmark points were detected: mitral valve hinge points at the intersection between the left atrium and the LV on the 2CH and 4CH LAX images, and right ventricular (RV) free wall insertion points at the intersection between the RV free wall and the interventricular septum on the SAX images. Mitral valve points were used to model the basal extent of the LV model, whereas RV insertion points were used to model the position of the septum. Inner (endocardial) and outer (epicardial) boundaries of the LV were defined on the SAX images. A finite element shape model was customized to the landmark and contour data, correcting for inter-slice breath-hold misregistration. The following sections describe each part of the pipeline. Details of the training and test datasets used in this study are shown in Figure 2.

## 2.3 Landmark Detection

Details of the landmark detection architecture, including training schemes, are given in Appendix A. Although sharing the same architecture, we trained three separate landmark detection networks to detect the different types of cardiac landmark points and image views: 2CH mitral valve points, 4CH mitral valve points and SAX RV insert points. A transfer learning training scheme was employed (see Appendix A) in which views were presented in random order until convergence.

In total, 2,379 participant CMR examinations with adequate LV landmark annotations were included (Table 1, Figure 2). Two experienced CMR analysts manually annotated the landmarks using the Cardiac Image Modeller software (version 6.2; Auckland MR Image Research Group, University of Auckland, New Zealand). Figure 4 shows examples of annotated landmark points on each image view. On average, five frames per case were annotated for the mitral valve points on each of the 4CH and 2CH views, and five short-axis slices per case were annotated for the RV inserts on the end-diastolic frame. These points resulted in 11,604 annotated frames for the two-chamber view, 11,670 for the four-chamber view, and 13,402 slices for the short-axis view.

Of the 2,379 exams, 50 were randomly selected for the test dataset, for which both CMR analysts annotated each exam to obtain an estimate of inter-observer error. The remaining 2,329 exams were used to train the landmark detection network. These exams were divided into training and validation sets on an exam-wise basis. The division was 2,097 (90%) cases for training and 232 (10%) cases for validation to monitor overfitting during training. Images were whitened by subtracting the mean pixel intensity and divided by standard deviation, on a per-image basis. Zero-padded cropping was performed to create 256x256 input images as needed.

We validated the predicted landmark points by the Euclidean distance (in mm) on the image space. The strength of agreement between the landmark detection and the two observers was measured using the intraclass correlation coefficient (ICC) with a two-way random effects model (17). A high ICC (close to 1) indicates a high similarity between landmark point locations from all observers.

## 2.4 Segmentation

Details of the segmentation architecture, including training schemes, are given in Appendix B. We used a U-Net architecture (16) to predict myocardial and cavity masks from short-axis images. Each image was segmented separately; no temporal or other spatial multi-slice information was learned for this segmentation network. During training, data augmentation was performed by image flipping, zoom, brightness, and contrast variations.

We included 1,545 exams where manual contours from the MESA CMR Core Lab (Johns Hopkins Medical Center, Baltimore, USA) were available (Table 1, Figure 2). The Core Lab analysis protocol for MESA study has been described previously (13). Briefly, endocardial and epicardial borders were traced by trained technologists on short-axis slices at end-diastole (ED) and end-systole (ES) frames using Q-MASS software (version 4.2, Medis, the Netherlands). Papillary muscles were included in the blood pool. All image contours were reviewed and corrected by a cardiac MR physician. In the core lab contour database, there were no links between contours and 3D image positions. Hence, we applied a simple algorithm to match and align contours with images. This consisted of ordering the images and contours from apex and base then alignment based on image plane and orientation. The alignment results were manually reviewed and mis-aligned cases were rejected (Figure 2).

The 1,545 exams were randomly split into 80% training cases (19,631 images), 10% validation cases (2,487 images) and 10% test cases (2,465 images). Masks images were generated from the epicardial and endocardial contours, providing three areas: myocardium, LV cavity, and background pixels. Images were zero-padded and cropped into 256x256 image size as needed.

We validated the accuracy of the segmentation network by using the Dice score (18), for both myocardium or LV cavity. We also validated standard clinical measurements for post-processing CMR exams (1), which include LV volumes at end-diastole and end-systole, ejection fraction and LV mass. Volumes were estimated by the LV cavity areas times the slice thickness (and slice gaps) for all short-axis slices where endocardial contours were available. LV masses were calculated from the myocardial volume (defined between epicardial and epicardial contours) multiplied by 1.05 g/ml constant. All volumes and masses were indexed by body surface area, resulted in LV end-diastolic volume index (LVEDVi), LV end-systolic volume index (LVESVi), LV mass index (LVMi). Ejection fraction (LVEF) was measured by (LVEDVi – LVESVi) / LVEDVi * 100. We compared all these values from the test cases (n=155) using the Bland-Altman plot analysis (19) to identify if there is a systematic error from the mean offset of the differences, inconsistent variability from the limits of agreement (mean ± 1.96 x standard deviation), and any trend of proportional error.

## 2.5 Atlas Construction

After landmark detection and segmentation (Figure 1), a finite element LV model was automatically customized to each set of myocardial contours and landmark points, as described previously (20). First, the LV model was fitted to the landmark and contour points by a least squares optimization. The extent of the LV was defined from landmarks on mitral valve points and an LV apex point obtained from the contours. The septum area was located using the RV insertion landmark points. After orienting the model according to the landmarks, the endocardial and epicardial surfaces were fitted to the short axis contours by minimizing the distance between the surfaces and the contour points. Mis-registrations of the contours due to differences in the breath-hold position from slice to slice were automatically corrected by shifting the contours in-plane to match an initial highly regularized model fit.

An LV atlas could be constructed by concatenating LV models from ED and ES frames to capture both shape and motion information (7). Procrustes alignment (21) was applied to the LV model to remove variations due to position and orientation. After alignment, the mean shape was calculated and principal component analysis (PCA) was applied to the registered LV shape models.

To demonstrate the clinical efficacy of the predicted LV atlas, we analyzed associations between LV shape and cardiovascular risk factors, i.e. hypertension, diabetes, smoking status, cholesterol level, and calcium score, and compared atlas associations obtained from the automatic pipeline with atlas associations obtained from manual contours and landmarks. For this evaluation, we evaluated 1,052 MESA cases independent of the sub-cohorts used to train the landmark and segmentation networks (atlas validation dataset, Table 1 and Figure 2). Our hypothesis was that there is no significant differences in the strength of risk factor associations between the automatically generated LV atlas and the atlas derived from manual analyses. Logistic regression (LR) models were used to evaluate the strength of the risk factor associations. A separate LR model was generated for each risk factor using that factor as a binary univariate dependent variable and the first 20 principal component scores derived from the atlas as the independent variables. The strength of the association between shape and risk factor was quantified using the area under the curve of the receiver operating characteristic (AUC). To avoid overfitting, a ten-fold cross validation scheme was employed.

# 3 Results

## 3.1 Landmark Detection

The total training time for three landmark detection networks was 14 hours on NVidia Titan X Pascal GPU. Typically, five iterations of transfer learning between 2CH, 4CH and SAX networks were required for overall convergence. The performance of the landmark detection networks was tested on 50 independent cases, which were annotated by two expert analysts independently. Only images where both analysts identified all landmark points were included. These resulted in 111 2CH, 107 4CH, and 286 SAX images for comparisons. Since two points are identified from each image, the total number of points during the test was 222, 214 and 572 points for 2CH, 4CH, and SAX respectively.

The distributions of Euclidean distances between automated methods and the observers are shown in Figure 3. Mean, standard deviation, and maximum distances are given in Table 2. The results show that the automated landmark detection errors are within the inter-observer variabilities with no significant differences in the location of landmark points (all $p<0.001$). ICC between the automated method and the two analysts were all excellent, i.e., 0.998, 0.996, and 0.995 for 2CH, 4CH, and SAX respectively.

Examples of landmark detections are shown in Figure 4 together with manual expert observer placements. The top row images show the largest distance of the automated detection method where the distance between observers was low (< 3 pixels). Even in these cases, the automated method could identify the landmarks very close to the observers. The bottom row images in Figure 4 showcase the largest distances between expert observers. The automated method was able to identify landmark points in these cases with the position very close to one of the observers. These cases show the difficulty of visually identifying landmark points where image contrast is low and high image noise is present.

### 3.2 Segmentation

Quartiles, means, and standard deviations of the Dice score from the test dataset are presented in

Table 3. Median and mean Dice scores were high (>0.8) for myocardium and LV cavity masks, both at ED and ES frames. Typical segmentation results are shown in Figure 5 with cases of best, mean, and worst results. Figure 5 also demonstrates the difficulty of segmenting basal slices near the LV outflow tract.

Table 4 shows comparisons of volumes (LVEDVi and LVESVi), mass (LVMi) and ejection fraction (LVEF) from the test cases. The segmentation network achieved excellent correlation coefficients for all clinical measurements (all Pearson's coefficients are > 0.9, $p<0.001$). The mean offset of differences are also small, i.e., less than 1 ml/$m^2$ for volumes, only 0.7% for ejection fraction, and 3 g/$m^2$ for mass. As shown in Figure 6, the differences are consistent within the limit of agreement lines without any visible trend for proportional error.

### 3.3 Atlas validation

Finally, we compared cardiovascular risk factor associations from the LV atlas from the automated analysis pipeline with an atlas formed from the manual analyses using a similar analysis method to (20). Table 5 shows the comparison of the area under the receiver operating characteristic curves (AUC) from risk factor association results (test cases from the cross validation). From all risk factors (hypertension, diabetes, smoking status, cholesterol, and calcium score), none of them have significant differences between the two methods except for cholesterol ($p=0.02$) which showed a stronger association with the automated analysis than with the manual analysis.

## 4 Discussion

In this study, we present methods for the automated analysis of large cohort data from a legacy dataset obtained in the MESA study, aided by deep learning methods. These methods enable a more complete analysis of large cohort datasets, augmenting the parameter set available from these valuable studies. In addition to the end-diastolic and end-systolic volumes computed in the original study, these methods enable the analysis of 3D shapes, facilitating a fully automated 3D model-based atlas analysis method. Almost all risk factors showed similar strength of relationships with atlas scores, except for cholesterol level in which the automated method showed a stronger relationship (Table 5). This indicates that the automated atlas analysis pipeline can be used in future studies in place of manual analyses.

The automated landmark detection method was successfully applied to the legacy gradient echo images, which are known to have lower signal-to-noise ratio and lower contrast compared to the current standard steady state free precession CMR imaging methods (5). The agreements with two expert analysts were all excellent (ICC > 0.9). Since signal-to-noise ratio is low in some gradient echo images, the analysts had noticeable disagreements between them in some cases, as shown in Figure 4 (bottom row). However, the automated detection method could identify the location of the landmark point in agreement with one of the observers. This ability was achieved by our approach to transfer learning weight parameters between image views iteratively. We exploited features between different domains to make the detection robust to noise and other artifacts.

Other machine learning methods have reported good results with landmark detection in cardiac MRI data, as well. For instance, Tarroni *et al.* (22) applied a hybrid random forest approach integrating both regression and structured classification models and reported mean errors of 3.2-3.9 mm in mitral valve landmark detection. Although it is difficult to determine which methods give the "best performance" in this application, our results show that the CNN-based method is powerful enough in the applications where legacy datasets provide sufficient annotated cases.

For the segmentation task, we demonstrated that the popular U-Net architecture (16) without any major modifications is capable of providing acceptable segmentation of the myocardium in gradient echo cine images. Although basal slices were more difficult for the network, the atlas shape customization method was relatively robust to segmentation mask outliers, as evidenced by the agreement in statistical relationships with common risk factors, since the model customization process used data from all slices.

It is known that different groups annotate cardiac MRI data differently (23). For this study, the manual contours were performed by a single core lab, whereas the landmarks were performed in another core lab, so both the landmark detection and segmentation networks will reflect the core lab standard operating procedures on the gradient echo images. Differences in local shape are expected when comparing the shape models generated with gradient echo imaging with those generated from other protocols, and these can be corrected using atlas-based methods (24). Alternatively, the training data distribution can be made richer to include more pathologies, images from different centers and multiple observers, as has been demonstrated by Tao *et al*. (25) and Bhuva *et al*. (26).

A common approach to train a complex deep learning network is by end-to-end training (27,28), where a combined loss function is defined for multiple tasks as the global cost function to optimize. In this work, landmarks and contours were only available on separated image views, so we decided to train the landmark detection network separately to the segmentation network to make each network capable of predicting unseen images independently. The ability to identify mitral valve points therefore does not need to depend on the segmentation masks or vice versa.

The problem of missing information is common to legacy datasets such as MESA. In this study, information linking contours with the corresponding 3D image position was not available. Since most cases were able to be matched with a simple algorithm, leading to sufficient training data, we did not invest more time in developing more sophisticated image-contour matching algorithms. This could be done in future work if an application requires larger training datasets.

## 5 Conclusions and Future Work

We have shown that deep learning networks can be used for automatically finding LV landmarks and segmentations on legacy MESA CMR images, in order to automate the construction of LV models, which can be used to build an atlas and evaluate associations between LV shape and risk factors. The final prediction of the LV model based on deep learning networks had similar power to evaluate associations with cardiovascular risk factors compared to manual analysis. This has greatly reduced the amount of time to analyze large-scale collections of cardiac MRI study. In future work, the automated atlas will be used to derive associations between LV shape and outcomes. In addition, analysis of all frames in the cine will allow the calculation of ejection and filling rates and other dynamic information.

## 6 Disclosures

The authors declare that the research was conducted in the absence of any commercial or financial relationships that could be construed as a potential conflict of interest. The views expressed in this manuscript are those of the authors and do not necessarily represent the views of the National Heart,

Lung, and Blood Institute; the National Institutes of Health; or the U.S. Department of Health and Human Services.

## 7 Author Contributions

All authors participated in analysis and interpretation of data, drafting of the manuscript, revising it critically, and final approval of the submitted manuscript.

## 8 Funding

This research was funded by the Health Research Council of New Zealand (17/608 and 17/234).

## 9 Acknowledgments

We would like to thank Benjamin Wen and Augustin Okamura for their analysis of the data. We also acknowledge the support of NVIDIA Corporation with the donation of the Titan X Pascal GPU and the Centre for eResearch at the University of Auckland for facilitating a GPU virtual machine used for this research. MESA and the MESA SHARe project are conducted and supported by the National Heart, Lung, and Blood Institute (NHLBI) in collaboration with MESA investigators. Support for MESA is provided by contracts N01-HC- 95159, N01-HC-95160, N01-HC-95161, N01-HC-95162, N01-HC-95163, N01-HC-95164, N01-HC-95165, N01-HC- 95166, N01-HC-95167, N01-HC-95168, N01-HC-95169 and CTSA UL1-RR-024156.

## 10 Data Availability Statement

The datasets analyzed for this study are available on request from the Cardiac Atlas Project (www.cardiacatlas.org).

## References

1. Schulz-Menger J, Bluemke DA, Bremerich J, Flamm SD, Fogel MA, Friedrich MG, Kim RJ, von Knobelsdorff-Brenkenhoff F, Kramer CM, Pennell DJ, et al. Standardized image interpretation and post-processing in cardiovascular magnetic resonance - 2020 update: Society for Cardiovascular Magnetic Resonance (SCMR): Board of Trustees Task Force on Standardized Post-Processing. *J Cardiovasc Magn Reson* (2020) 22: doi:10.1186/s12968-020-00610-6

2. Bild DE. Multi-Ethnic Study of Atherosclerosis: Objectives and Design. *Am J Epidemiol* (2002) 156:871–881. doi:10.1093/aje/kwf113

3. Petersen SE, Matthews PM, Bamberg F, Bluemke DA, Francis JM, Friedrich MG, Leeson P, Nagel E, Plein S, Rademakers FE, et al. Imaging in population science: cardiovascular magnetic resonance in 100,000 participants of UK Biobank - rationale, challenges and approaches. *J Cardiovasc Magn Reson* (2013) 15:46. doi:10.1186/1532-429X-15-46

4. Kramer CM, Barkhausen J, Bucciarelli-Ducci C, Flamm SD, Kim RJ, Nagel E. Standardized cardiovascular magnetic resonance imaging (CMR) protocols: 2020 update. *J Cardiovasc Magn Reson* (2020) 22: doi:10.1186/s12968-020-00607-1

5. Malayeri AA, Johnson WC, Macedo R, Bathon J, Lima JAC, Bluemke DA. Cardiac cine MRI: Quantification of the relationship between fast gradient echo and steady-state free precession for determination of myocardial mass and volumes. *J Magn Reson Imaging JMRI* (2008) 28:60–66. doi:10.1002/jmri.21405

6. Medrano-Gracia P, Cowan BR, Ambale-Venkatesh B, Bluemke DA, Eng J, Finn JP, Fonseca CG, Lima JA, Suinesiaputra A, Young AA. Left ventricular shape variation in asymptomatic populations: the multi-ethnic study of atherosclerosis. *J Cardiovasc Magn Reson* (2014) 16: doi:10.1186/s12968-014-0056-2

7. Zhang X, Cowan BR, Bluemke DA, Finn JP, Fonseca CG, Kadish AH, Lee DC, Lima JAC, Suinesiaputra A, Young AA, et al. Atlas-Based Quantification of Cardiac Remodeling Due to Myocardial Infarction. *PLoS ONE* (2014) 9:e110243. doi:10.1371/journal.pone.0110243

8. Suinesiaputra A, Dhooge J, Duchateau N, Ehrhardt J, Frangi AF, Gooya A, Grau V, Lekadir K, Lu A, Mukhopadhyay A, et al. Statistical Shape Modeling of the Left Ventricle: Myocardial Infarct Classification Challenge. *IEEE J Biomed Health Inform* (2018) 22:503–515. doi:10.1109/JBHI.2017.2652449

9. Piras P, Teresi L, Puddu PE, Torromeo C, Young AA, Suinesiaputra A, Medrano-Gracia P. Morphologically normalized left ventricular motion indicators from MRI feature tracking characterize myocardial infarction. *Sci Rep* (2017) 7:12259. doi:10.1038/s41598-017-12539-5

10. Albà X, Lekadir K, Pereañez M, Medrano-Gracia P, Young AA, Frangi AF. Automatic initialization and quality control of large-scale cardiac MRI segmentations. *Med Image Anal* (2018) 43:129–141. doi:10.1016/j.media.2017.10.001

11. Leiner T, Rueckert D, Suinesiaputra A, Baeßler B, Nezafat R, Išgum I, Young AA. Machine learning in cardiovascular magnetic resonance: basic concepts and applications. *J Cardiovasc Magn Reson* (2019) 21: doi:10.1186/s12968-019-0575-y

12. Litjens G, Kooi T, Bejnordi BE, Setio AAA, Ciompi F, Ghafoorian M, van der Laak JAWM, van Ginneken B, Sánchez CI. A survey on deep learning in medical image analysis. *Med Image Anal* (2017) 42:60–88. doi:10.1016/j.media.2017.07.005

13. Natori S, Lai S, Finn JP, Gomes AS, Hundley WG, Jerosch-Herold M, Pearson G, Sinha S, Arai A, Lima JAC, et al. Cardiovascular Function in Multi-Ethnic Study of Atherosclerosis: Normal Values by Age, Sex, and Ethnicity. *Am J Roentgenol* (2006) 186:S357–S365. doi:10.2214/AJR.04.1868

14. Fonseca CG, Backhaus M, Bluemke DA, Britten RD, Chung JD, Cowan BR, Dinov ID, Finn JP, Hunter PJ, Kadish AH, et al. The Cardiac Atlas Project—an imaging database for computational modeling and statistical atlases of the heart. *Bioinformatics* (2011) 27:2288–2295. doi:10.1093/bioinformatics/btr360

15. Simonyan K, Zisserman A. Very Deep Convolutional Networks for Large-Scale Image Recognition. in *3rd International Conference on Learning Representations, ICLR 2015, San Diego, CA, USA, May 7-9, 2015, Conference Track Proceedings*, eds. Y. Bengio, Y. LeCun Available at: http://arxiv.org/abs/1409.1556

16. Ronneberger O, Fischer P, Brox T. "U-Net: Convolutional Networks for Biomedical Image Segmentation," in *Medical Image Computing and Computer-Assisted Intervention – MICCAI 2015*, eds. N. Navab, J. Hornegger, W. M. Wells, A. F. Frangi (Cham: Springer International Publishing), 234–241. doi:10.1007/978-3-319-24574-4_28

17. Koo TK, Li MY. A Guideline of Selecting and Reporting Intraclass Correlation Coefficients for Reliability Research. *J Chiropr Med* (2016) 15:155–163. doi:10.1016/j.jcm.2016.02.012

18. Eelbode T, Bertels J, Berman M, Vandermeulen D, Maes F, Bisschops R, Blaschko MB. Optimization for Medical Image Segmentation: Theory and Practice When Evaluating With Dice Score or Jaccard Index. *IEEE Trans Med Imaging* (2020) 39:3679–3690. doi:10.1109/TMI.2020.3002417

19. Martin Bland J, Altman DouglasG. Statistical Methods for Assessing Agreement Between Two Methods of Clinical Measurement. *The Lancet* (1986) 327:307–310. doi:10.1016/S0140-6736(86)90837-8

20. Gilbert K, Bai W, Mauger C, Medrano-Gracia P, Suinesiaputra A, Lee AM, Sanghvi MM, Aung N, Piechnik SK, Neubauer S, et al. Independent Left Ventricular Morphometric Atlases Show Consistent Relationships with Cardiovascular Risk Factors: A UK Biobank Study. *Sci Rep* (2019) 9:1130. doi:10.1038/s41598-018-37916-6

21. Dryden IL, Mardia KV. *Statistical shape analysis*. Chichester ; New York: John Wiley & Sons (1998).

22. Tarroni G, Bai W, Oktay O, Schuh A, Suzuki H, Glocker B, Matthews PM, Rueckert D. Large-scale Quality Control of Cardiac Imaging in Population Studies: Application to UK Biobank. *Sci Rep* (2020) 10:2408. doi:10.1038/s41598-020-58212-2

23. Suinesiaputra A, Bluemke DA, Cowan BR, Friedrich MG, Kramer CM, Kwong R, Plein S, Schulz-Menger J, Westenberg JJM, Young AA, et al. Quantification of LV function and mass by cardiovascular magnetic resonance: multi-center variability and consensus contours. *J Cardiovasc Magn Reson* (2015) 17: doi:10.1186/s12968-015-0170-9

24. Medrano-Gracia P, Cowan BR, Bluemke DA, Finn JP, Kadish AH, Lee DC, Lima JAC, Suinesiaputra A, Young AA. Atlas-based analysis of cardiac shape and function: correction of regional shape bias due to imaging protocol for population studies. *J Cardiovasc Magn Reson* (2013) 15:80. doi:10.1186/1532-429X-15-80

25. Tao Q, Yan W, Wang Y, Paiman EHM, Shamonin DP, Garg P, Plein S, Huang L, Xia L, Sramko M, et al. Deep Learning–based Method for Fully Automatic Quantification of Left Ventricle Function from Cine MR Images: A Multivendor, Multicenter Study. *Radiology* (2019) 290:81–88. doi:10.1148/radiol.2018180513

26. Bhuva AN, Bai W, Lau C, Davies RH, Ye Y, Bulluck H, McAlindon E, Culotta V, Swoboda PP, Captur G, et al. A Multicenter, Scan-Rescan, Human and Machine Learning CMR Study to Test Generalizability and Precision in Imaging Biomarker Analysis. *Circ Cardiovasc Imaging* (2019) 12: doi:10.1161/CIRCIMAGING.119.009214

27. Song L, Lin J, Wang ZJ, Wang H. An End-to-End Multi-Task Deep Learning Framework for Skin Lesion Analysis. *IEEE J Biomed Health Inform* (2020) 24:2912–2921. doi:10.1109/JBHI.2020.2973614

28. Yap MH, Goyal M, Osman FM, Martí R, Denton E, Juette A, Zwiggelaar R. Breast ultrasound lesions recognition: end-to-end deep learning approaches. *J Med Imaging Bellingham Wash* (2019) 6:011007. doi:10.1117/1.JMI.6.1.011007

**Figures / Tables:**

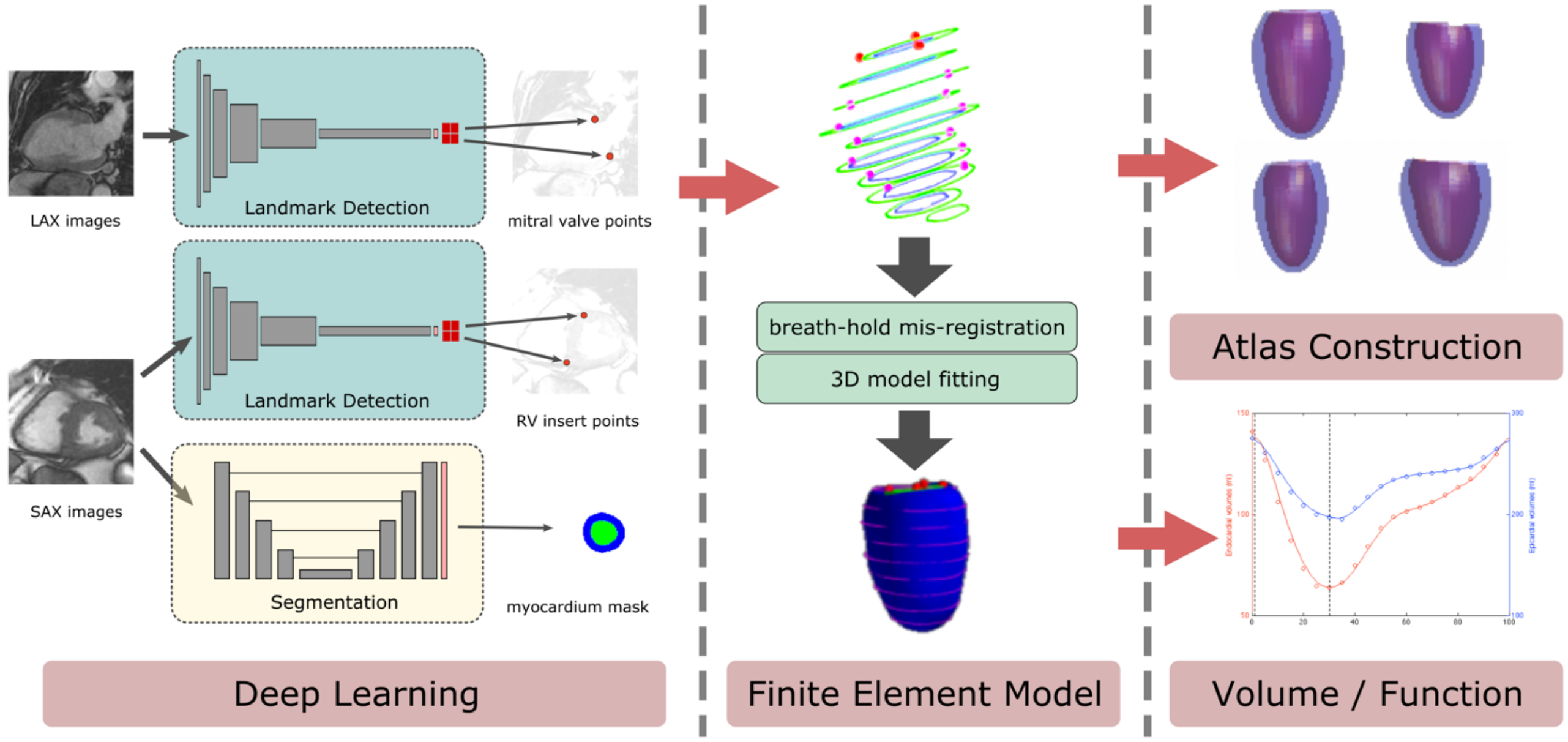

**Figure 1**. Fully-automated atlas generation pipeline of cardiac MRI analyses. Three deep learning networks were trained to perform: 1) detection of mitral valve points from long-axis (LAX) images, from both two-chamber or four-chamber views, 2) detection of right ventricular (RV) insert points from short-axis (SAX) images, and 3) segmentation of myocardium mask from SAX images. Landmark points and contours from myocardium mask images were converted into 3D patient coordinates to guide the customization of a left ventricle (LV) model. Breath-hold mis-registration of SAX slices were corrected. The final model was used to construct a statistical shape LV atlas.

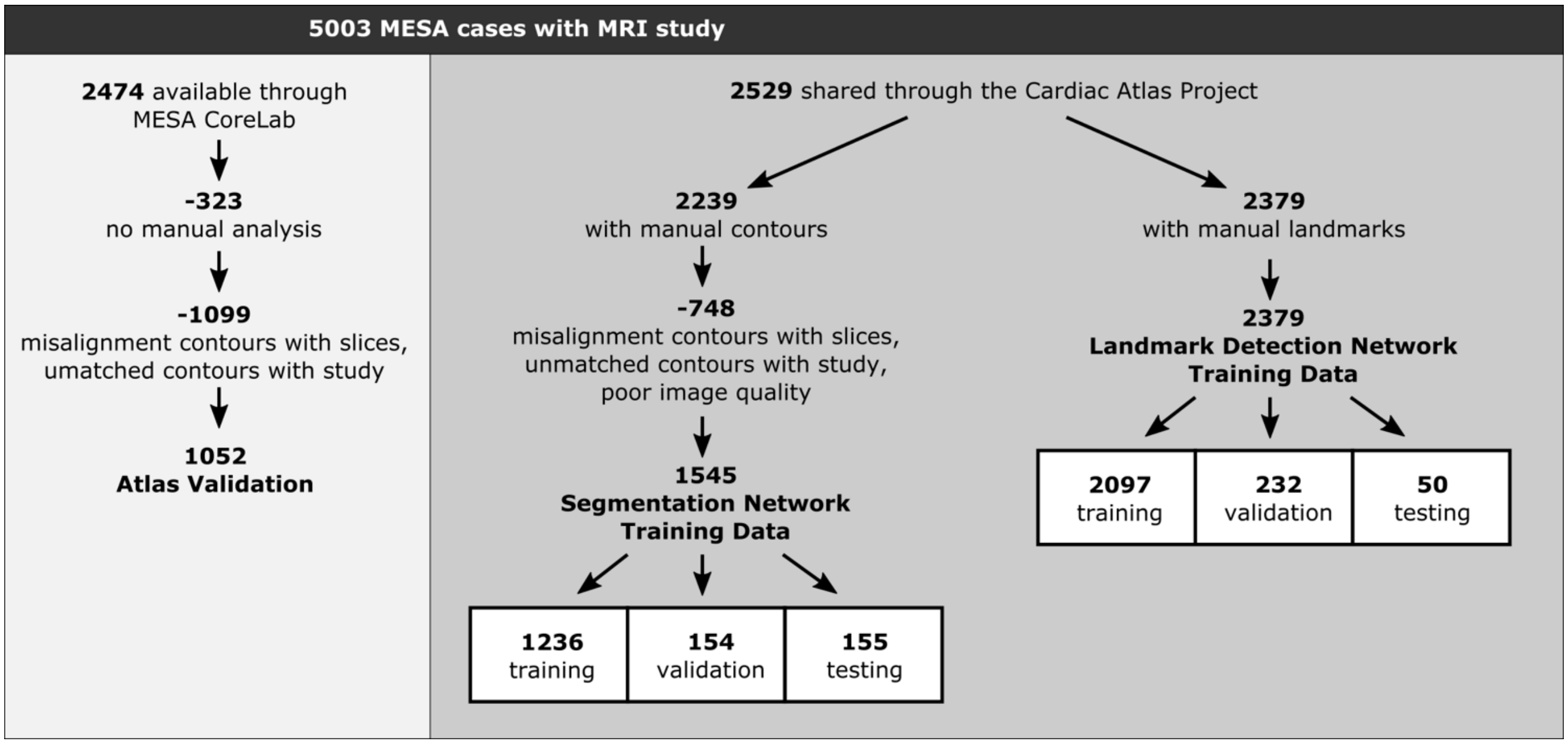

**Figure** 2. Division of MESA cases into two independent sets of Atlas Validation and Training sub-cohorts. Within the Training sub-cohort, cases were divided into training, validation and testing sub-groups for the different deep learning networks (Segmentation Network and Landmark Detection Network).

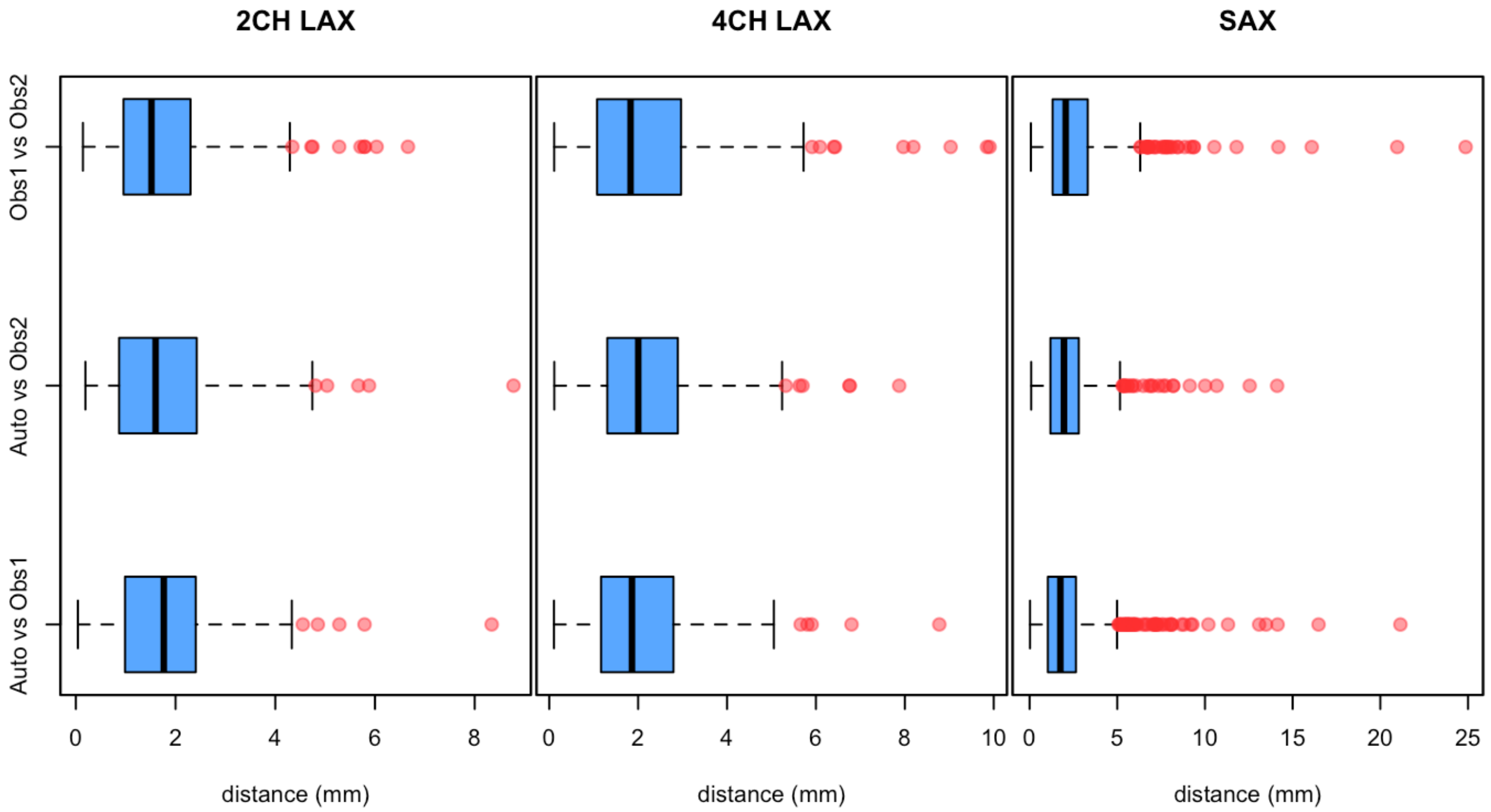

**Figure 3.** Distributions of distances between landmark points identified by the landmark detection method (Auto) and the two analysts (Obs1 and Obs2). Median (solid line), quartiles (thin lines) outliers (red points).

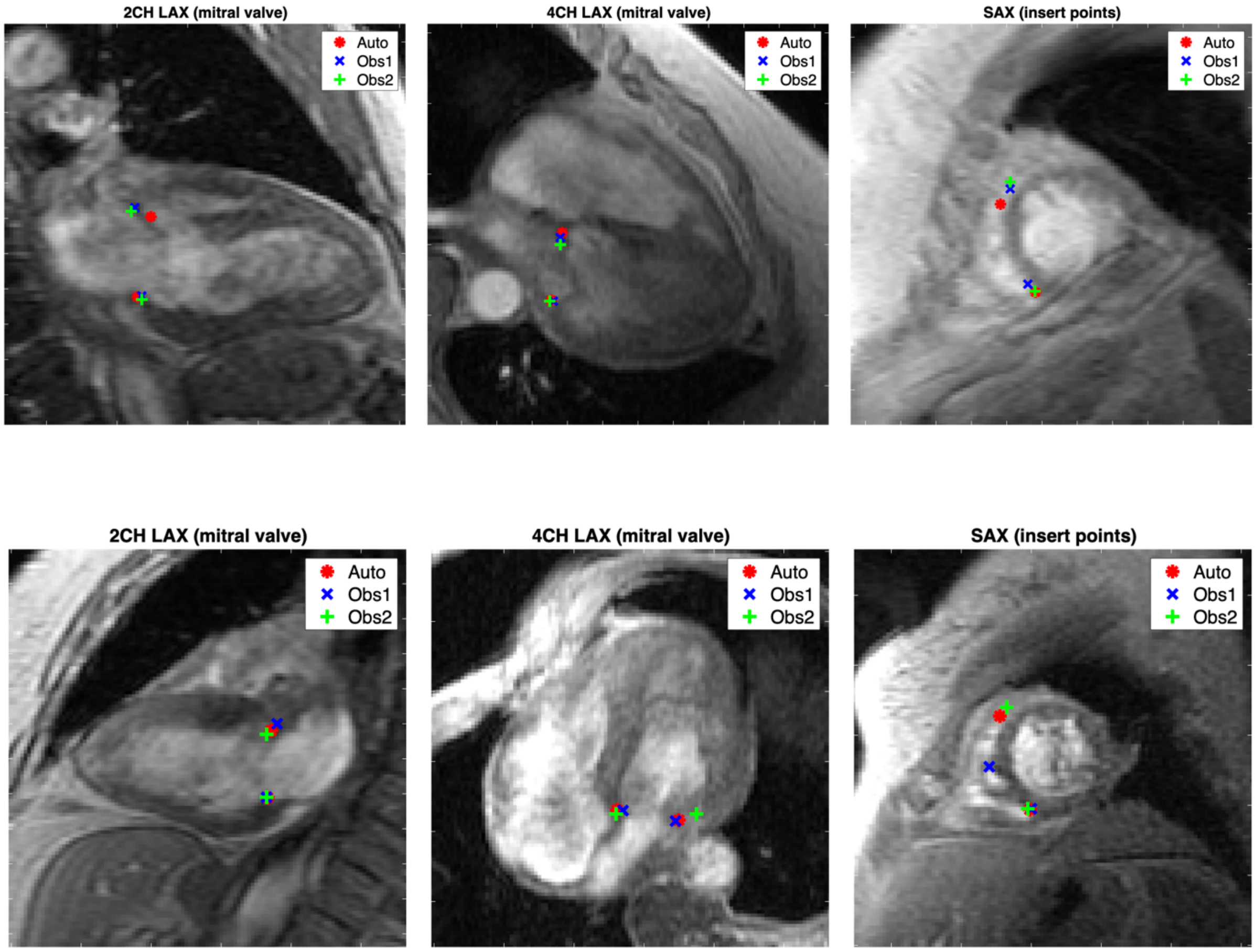

**Figure 4.** Examples of automated landmark detection (red markers) compared with manually defined placements by two observers (blue and green markers). The top row shows cases with the maximum distance of automated detection to one of the observers while interobserver distances are small. The bottom row shows cases with the largest interobserver distances.

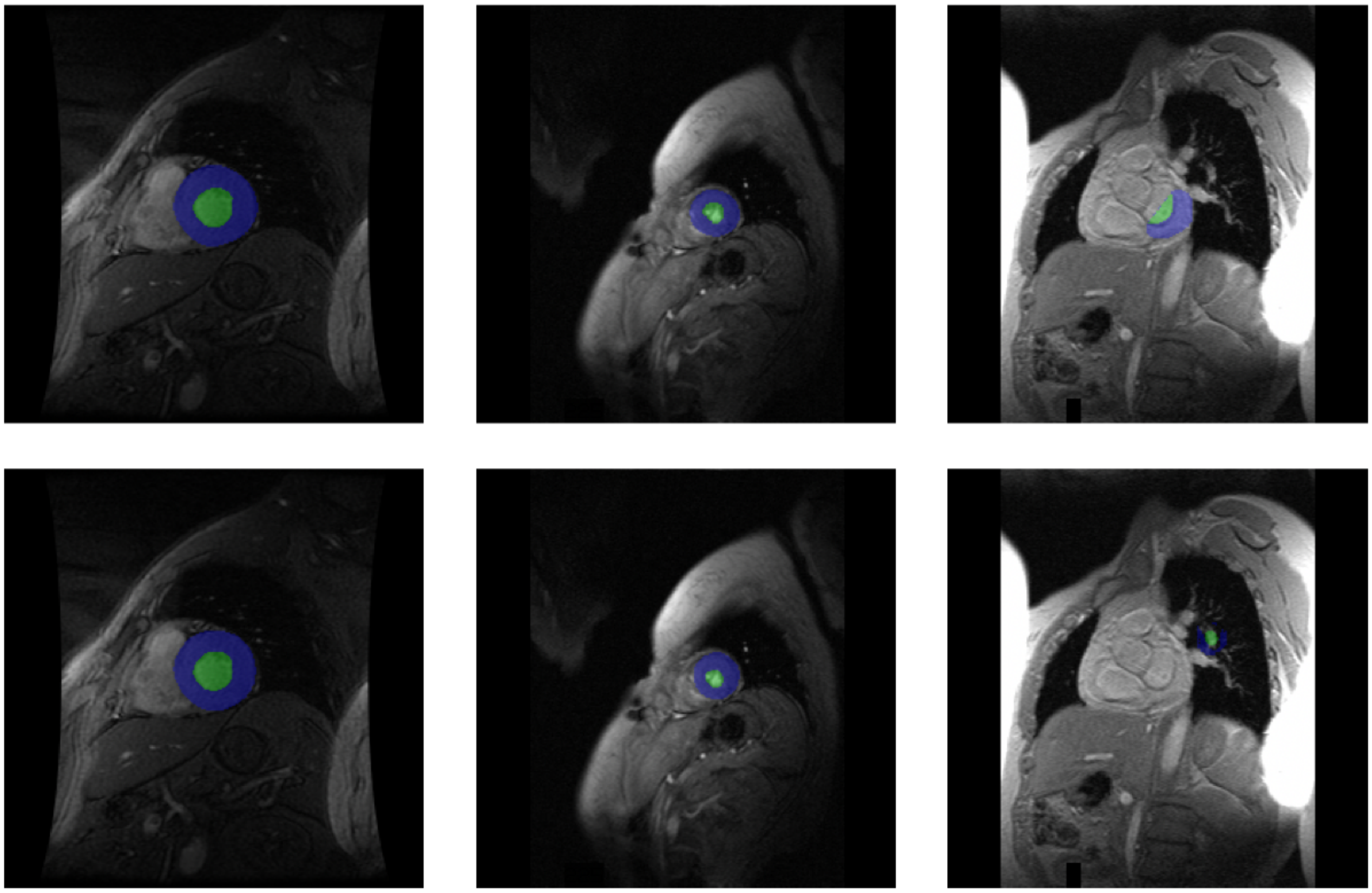

**Figure 5.** Examples of short axis segmentation network results. Top row: manual segmentation, bottom row: automated segmentation network. Left: predicted segmentations with the best Dice score, middle: a mean Dice score, right: the worst Dice score.

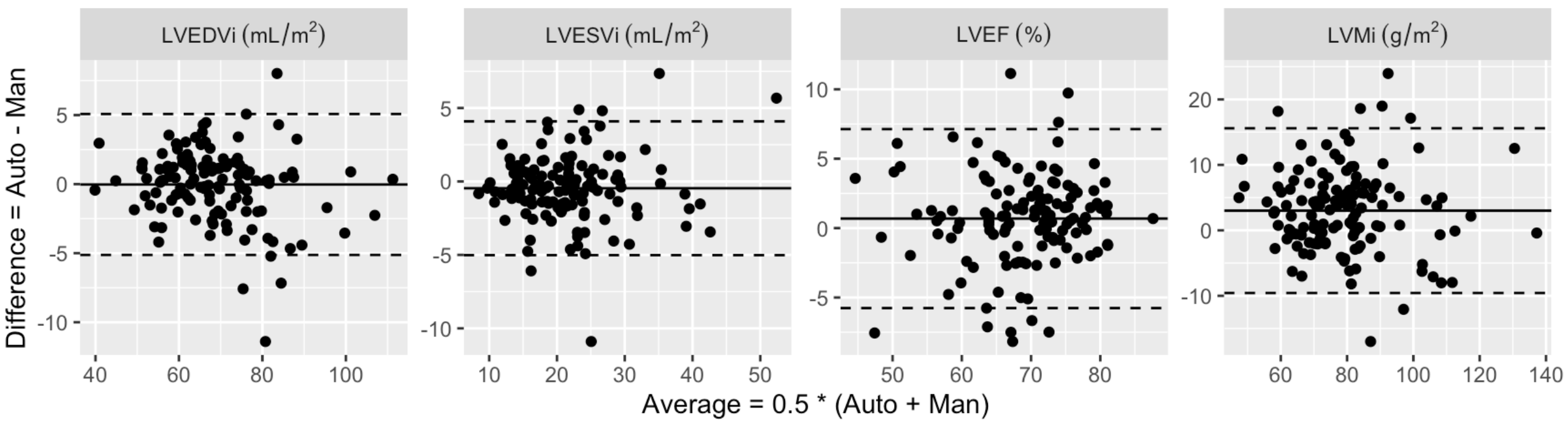

**Figure 6.** Differences between automated analysis (Auto) and manually drawn contours (Man). Solid lines are mean differences and dashed lines are the limits of agreement within ± 1.96 x standard deviation from the mean. The mean difference values are shown in Table 4.

**Table 1**. Patient demographics from the MESA cohort. Two sub-cohorts were defined to train and validate deep learning networks for landmark detection and segmentation. Another sub-cohort, disjoint from the two training datasets, was defined for validation of the atlas generated from automated compared with core lab manual analysis. Continuous variables are written as mean (standard deviation), while categorical variables are written as count (percentage). Statistical tests were performed between a sub-cohort against its complement with one-way ANOVA for continuous variables and $\chi^2$ test for categorical variables. * p<0.05, $p<0.01, #p<0.001 for difference between a particular sub-cohort and the rest of the MESA CMR cohort.

| | | MESACMR | Landmark Detection | Segmentation | Atlas Validation |
|---|---|---|---|---|---|
| N | | 5003 | 2379 | 1545 | 1052 |
| Age (years) | | 61.5 (10.1) | 61.3 (10.1) | 61.0 (10.2) $ | 60.1 (9.8) # |
| Gender | Female | 2622 (52.4) | 1213 (52.1) | 814 (52.7) | 430 (40.9) # |
| | Male | 2381 (47.6) | 1116 (47.9) | 731 (47.3) | 622 (59.1) |
| SBP (mmHg) | | 125.4(21.3) | 126.2 (21.9)* | 126.4 (22.0)* | 124.8 (20.2) |
| DBP (mmHg) | | 71.8 (10.30 | 71.6 (10.3) | 71.7 (10.3) | 73.6 (10.1) # |
| Heart Rate (bpm) | | 62.8 (9.5) | 62.7 (9.5) | 62.9 (9.5) | 62.1 (9.6) $ |
| Diabetes | Yes | 459 (9.2) | 228 (9.8) | 162 (10.5)* | 74 (7.0) $ |
| | No | 4544 (90.8) | 2101 (90.2) | 1383 (89.5) | 978 (93.0) |
| Hypertension | Yes | 1766 (35.3) | 791 (34.0) | 539 (34.9) | 373 (35.5) |
| | No | 3234 (64.7) | 1537 (66.0) | 1005 (65.1) | 677 (64.5) |
| Smoking Status | Never | 2569 (51.5) | 1222 (52.7) | 805 (52.4) | 511 (48.6) |
| | Former | 1786 (35.8) | 804 (34.7) | 521 (33.9) | 394 (37.5) |
| | Current | 634 (12.7) | 294 (12.7) | 209 (13.6) | 146 (13.9) |
| Framingham Score | | 13.9 (9.5) | 14.1 (9.5) | 14.0 (9.6) | 13.7 (9.2) |

**Table 2**. Differences and intraclass correlation (ICC) values in detecting landmarks on 50 validation cases. All difference values are expressed mean (standard deviation) from the Euclidean distance between annotations in millimeters.

| | 2CH LAX (n=222) | 4CH LAX (n=214) | SAX (n=572) |
|---|---|---|---|
| Auto vs Obs1 | 1.86 (1.19) | 2.09 (1.32) | 2.29 (2.15) |
| Auto vs Obs2 | 1.81 (1.21) | 2.19 (1.28) | 2.27 (1.61) |
| Obs1 vs Obs2 | 1.78 (1.16) | 2.24 (1.68) | 2.67 (2.29) |
| ICC value | 0.998 | 0.996 | 0.995 |

**Table 3.** Dice score results of the segmentation network from the test dataset (n=2,465 images). Frames indicate end-diastole (ED) and end-systole (ES). The 25th quartile (Q1), median, and 75th quartile (Q3) are shown, together with means and standard deviations.

| Mask | Frame | Q1 | Median | Q3 | Mean | Std Dev |
|---|---|---|---|---|---|---|
| Cavity | ED | 0.92 | 0.95 | 0.97 | 0.93 | 0.07 |
| | ES | 0.86 | 0.91 | 0.94 | 0.88 | 0.11 |
| Myocardium | ED | 0.85 | 0.89 | 0.91 | 0.87 | 0.07 |
| | ES | 0.89 | 0.92 | 0.94 | 0.90 | 0.08 |

**Table 4.** Comparisons of indexed LV volumes, ejection fraction and mass from the 155 test cases between the predicted segmentation results with manual contours. The differences are written as mean (standard deviation).

| LV Function | Correlation coefficient | Differences |
|---|---|---|
| LVEDVi (ml/m2) | 0.98 ($p<0.001$) | -0.02 (2.6) |
| LVESVi (ml/m2) | 0.95 ($p<0.001$) | -0.46 (2.3) |
| LVEF (%) | 0.92 ($p<0.001$) | 0.69 (3.3) |
| LVMi (g/m2) | 0.92 ($p<0.001$) | 3.0 (6.4) |

**Table 5.** Area under the ROC curve (AUC) comparisons from the 1052 LV shape association studies using different contours: manual (Man) and deep learning (Auto). *$p<0.05$ De Long test.

| | AUC | |
|---|---|---|
| | Man | Auto |
| Hypertension | 0.69 | 0.71 |
| Diabetes | 0.56 | 0.53 |
| Smoking status | 0.59 | 0.61 |
| Cholesterol | 0.50 | 0.54* |
| Calcium score | 0.61 | 0.61 |

**Appendixes**

## A. Landmark Detection Network

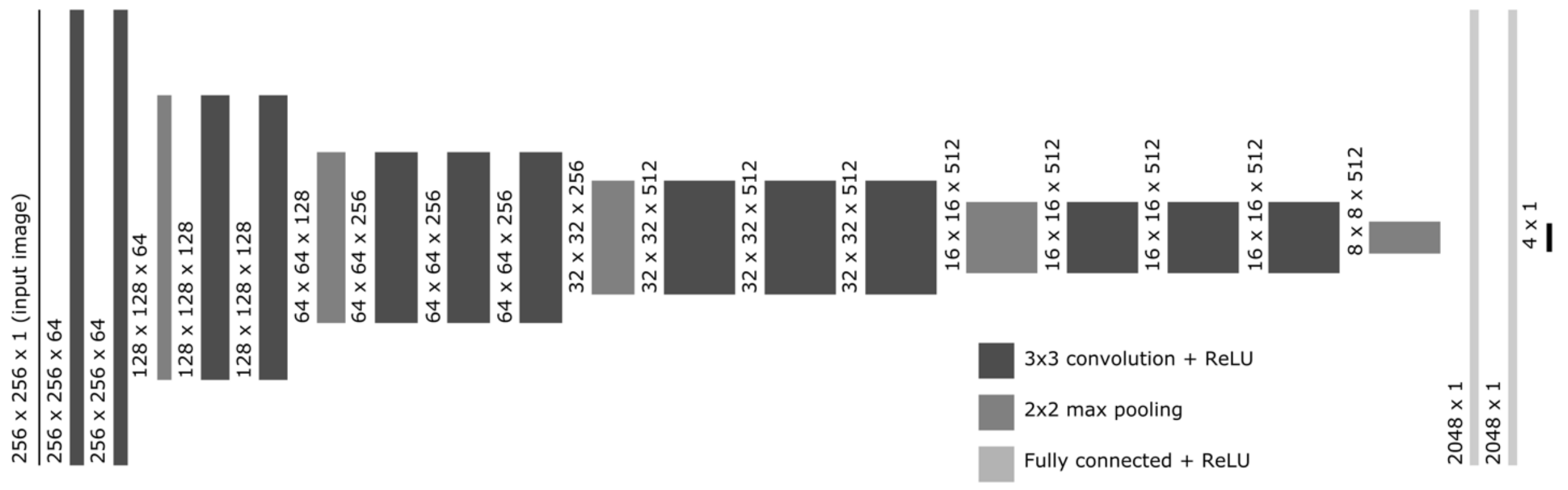

**Figure A-1.** Landmark detection deep learning network architecture based on VGG16.

The landmark detection network was based on the VGGNet architecture (15), which has been successfully used to classify images and to recognize objects. We used the 16-layer VGGNet and trained the network from scratch. We did not use a pre-trained network as a base, because of significant differences between CMR images compared to natural RGB images. We modified the network by reducing the number of neurons from 4096 to 2048 and the size of the final layer was changed to 4 neurons, corresponding to the two cardiac points to be annotated, i.e., pairs of (x, y) image coordinates. The network was trained using a uniform random initialization without temporal or other spatial information. Since two-chamber, four-chamber and short-axis views have distinctive appearances, we trained three separate networks independently, instead of sharing the weights.

The landmark detection network was implemented using Keras with the TensorFlow backend [20], AdaDelta optimizer for backpropagation, initial learning rate of 1.0, $E = 10^{-8}$ and $\rho = 0.95$, and a mini-batch size of 48. On stagnation of the validation loss for three consecutive epochs, the learning rate was reduced by a factor of 5. The network was deemed to be converged when the root mean square error of the normalized coordinates did not decrease by at least $10^{-6}$ in six consecutive epochs. With the exception of data-augmentation, no additional regularization was performed.

Although the three landmark detection networks were trained separately, we facilitated transfer learning between networks during training. This was designed to exploit similarities in the images, yet allowing for differences in the spatial relationships. First, an initial network for one view was trained from scratch with random weight initialization until convergence. Then, the network was retrained for one of the remaining two views. However, instead of using a random initialization, the weights from the previous training step were used as initial weights. After the new network was converged, its weights were used as initialization for the third view. The order in which the three different views were trained was random. This sequence was repeated until convergence (e.g. the performance of the two-chamber network compared to the previously trained two-chamber network was not improved). An advantage of this sequence is that it allows for maximum freedom when training the neural networks for the different kinds of image views whilst still being able to infer features learned from other images. Table A-1 shows the improvement of performance using the transfer learning scheme in the validation set, where the landmark distance errors decreased when the network was trained using our learning strategy.

To prevent overfitting and to increase generalization, data augmentation was applied at training time. Images were randomly rotated up to 45 degrees in either direction, shifted by up to 25 pixels in both $x$ and $y$ direction and sheared by a factor between $-0.2$ and $0.2$. The four-chamber and two-chamber were randomly flipped horizontally or vertically respectively, with a probability of 0.3, while the short axis images were randomly scaled with a factor between 0.8 and 1.2.

**Table A-1.** Landmark distance errors from neural networks trained independently compared to networks trained with our training strategy. Error values were measured on 232 validation cases and shown as mean (standard deviation). All values are in millimeters

| | System trained with independent neural networks | System trained with our training strategy |
|---|---|---|
| two-chamber | 2.98 (1.44) | 1.53 (0.74) |
| four-chamber | 3.24 (1.55) | 1.44 (0.74) |
| short axis | 2.94 (1.6) | 2.07 (1.11) |

## B. Segmentation Network

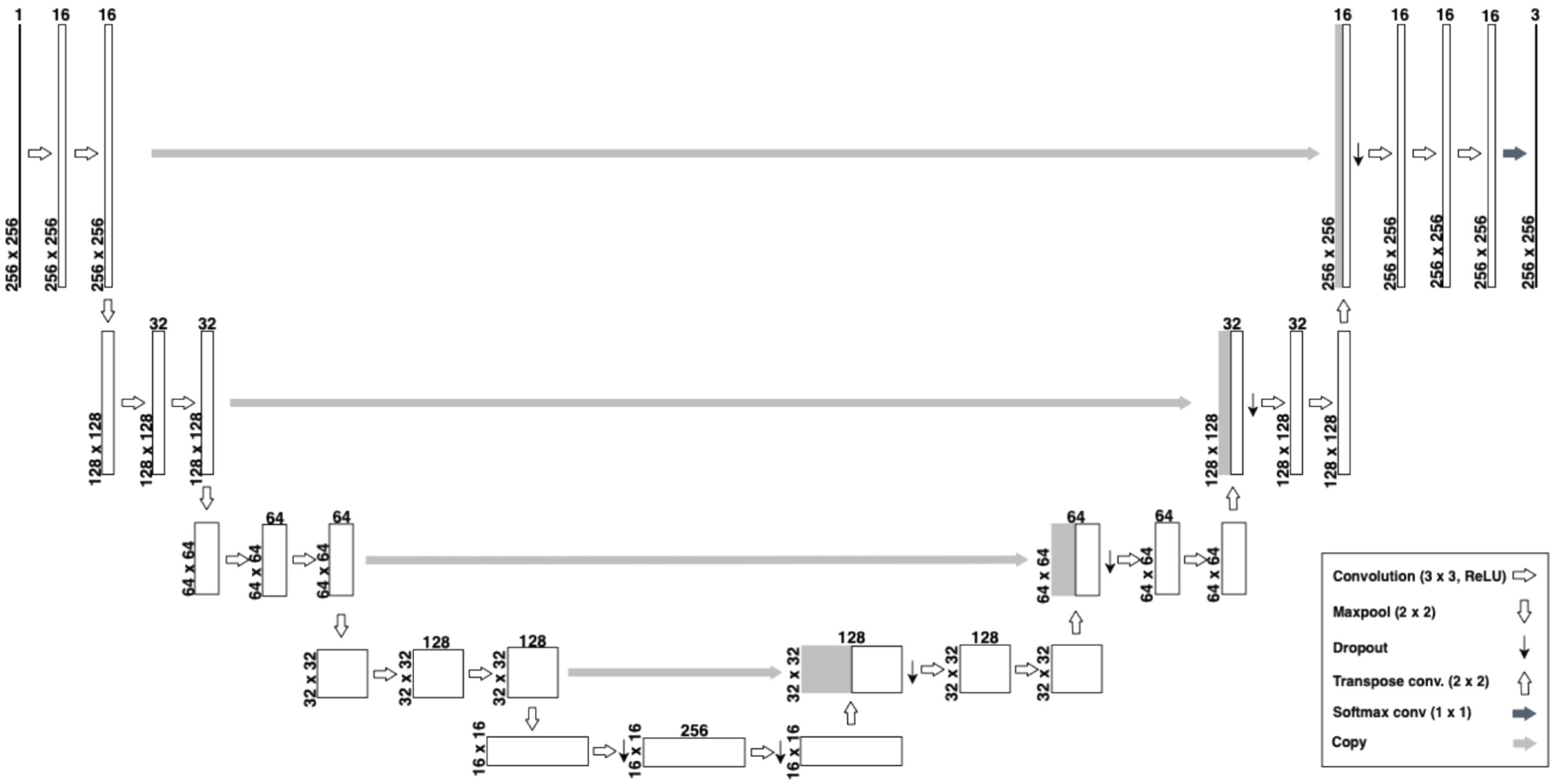

**Figure B-1.** Segmentation deep learning network architecture based on U-Net.

We used a U-Net architecture (16) to segment the myocardium. Briefly, the network consisted of a contracting path and an expansive path. The contracting path consisted of a repeated application of two 3×3 convolutions with batch normalization, each followed by zero-padding, a ReLU and a 2×2 max pooling operation with stride 2 for downsampling. Every step in the expansive path consisted of an upsampling of the feature map followed by a 2×2 convolution with batch normalization, followed by zero-padding and a ReLU. This was followed by a concatenation with the corresponding feature map from the contracting path, and two 3×3 convolutions with batch normalization, each followed by zero-padding and a ReLU. At the second last layer, a 3×3 convolution was performed followed by zero-padding and a ReLU, and at the final layer a 1×1 convolution was used to map each 16-component feature vector to the desired number of three classes. In total the network had 24 convolutional layers. The loss function was computed by a pixelwise softmax cost function over the final feature map combined with the dice loss function.

Images were normalized to have a zero mean and a unit standard deviation. Data augmentations were performed by rotation, random horizontal / vertical image flip, zoom, brightness, and contrast. For rotation augmentation, images were randomly rotated with an angle between 0 and 360 degrees. Bilinear interpolation was applied for the original MR images, while nearest neighbor interpolation was used for the mask images. For the zoom augmentation, the MR images were resized by a scalar of 3 to 8 using bicubic interpolation, and the same operation was applied for the image mask with nearest neighbor interpolation. The brightness/contrast augmentation randomly adjusted the brightness/contrast of the MR images by up to ±50%.

The initial weights of the layers in the segmentation network were random values, which were then optimized by stochastic gradient descent. The training optimization was dependent on the learning rate and loss function. The initial learning rate was 0.001 and was kept at this value for the first ten epochs, after which it was reduced by the square-root of two after every fifth epoch. The loss was calculated as 1 minus the weighted DSC for the myocardium and cavity, which was averaged for the slice.